\documentclass[10pt,conference]{IEEEtran}
\IEEEoverridecommandlockouts
\usepackage{cite}
\usepackage[numbers,sort&compress]{natbib}
\usepackage{amsmath,amssymb,amsfonts}
\usepackage{algorithmic}
\usepackage{graphicx}
\usepackage{textcomp}
\usepackage{xcolor}
\usepackage{float}
\usepackage{multirow} 
\usepackage{array}    
\usepackage{booktabs} 
\usepackage{afterpage}
\usepackage[caption=false,font=footnotesize]{subfig}

\usepackage{array} 
\usepackage[hidelinks]{hyperref}

\hypersetup{
    colorlinks=true,
    linkcolor=blue,      
    citecolor=green,     
    urlcolor=blue,       
    anchorcolor=blue     
}

\makeatletter
  \ifCLASSOPTIONconference
    \long\def\@makecaption#1#2{%
      \ifx\@captype\@IEEEtablestring%
        \footnotesize
        \bgroup\par\centering\@IEEEtabletopskipstrut{\normalfont\footnotesize {#1.} #2}\par\addvspace{0.5\baselineskip}\egroup%
        \@IEEEtablecaptionsepspace
      \else
        \@IEEEfigurecaptionsepspace
        \setbox\@tempboxa\hbox{\normalfont\footnotesize {#1.} #2}%
        \ifdim \wd\@tempboxa >\hsize%
          \setbox\@tempboxa\hbox{\normalfont\footnotesize {#1.}}%
          \parbox[t]{\hsize}{\normalfont\footnotesize \noindent\unhbox\@tempboxa#2}%
        \else
          \hbox to\hsize{\normalfont\footnotesize\hfil\box\@tempboxa\hfil}%
        \fi
      \fi}
  \fi
\makeatother

\def\BibTeX{{\rm B\kern-.05em{\sc i\kern-.025em b}\kern-.08em
    T\kern-.1667em\lower.7ex\hbox{E}\kern-.125emX}}
\begin{document}

\title{Quantum Annealing-Based Algorithm for Efficient Coalition Formation Among LEO Satellites}

\author{
    Supreeth Mysore Venkatesh\textsuperscript{1,2},
    Antonio Macaluso\textsuperscript{2},
    Marlon Nuske\textsuperscript{3},\\
    Matthias Klusch\textsuperscript{2}, 
    Andreas Dengel\textsuperscript{1,3}\\
    \textsuperscript{1}\textit{University of Kaiserslautern-Landau (RPTU)}, Kaiserslautern, Germany \\
    \textsuperscript{2}\textit{German Research Center for Artificial Intelligence (DFKI)}, Saarbruecken, Germany \\
    \textsuperscript{3}\textit{German Research Center for Artificial Intelligence (DFKI)}, Kaiserslautern, Germany \\
    {\texttt{supreeth.mysore@rptu.de}}, 
    {\texttt{antonio.macaluso@dfki.de}}, 
    {\texttt{marlon.nuske@dfki.de}},\\
    {\texttt{matthias.klusch@dfki.de}}, 
    {\texttt{andreas.dengel@dfki.de}}
}

\maketitle

\begin{abstract}

The increasing number of Low Earth Orbit (LEO) satellites, driven by lower manufacturing and launch costs, is proving invaluable for Earth observation missions and low-latency internet connectivity. However, as the number of satellites increases, the number of communication links to maintain also rises, making the management of this vast network increasingly challenging and highlighting the need for clustering satellites into efficient groups as a promising solution. This paper formulates the clustering of LEO satellites as a coalition structure generation (CSG) problem and leverages quantum annealing to solve it. We represent the satellite network as a graph and obtain the optimal partitions using a hybrid quantum-classical algorithm called GCS-Q. The algorithm follows a top-down approach by iteratively splitting the graph at each step using a quadratic unconstrained binary optimization (QUBO) formulation. To evaluate our approach, we utilize real-world three-line element set (TLE/3LE) data for Starlink satellites from Celestrak. Our experiments, conducted using the D-Wave Advantage annealer and the state-of-the-art solver Gurobi, demonstrate that the quantum annealer significantly outperforms classical methods in terms of runtime while maintaining the solution quality. The performance achieved with quantum annealers surpasses the capabilities of classical computers, highlighting the transformative potential of quantum computing in optimizing the management of large-scale satellite networks.

\end{abstract}


\section{Introduction}

In recent years, there has been a significant surge in interest and investment from various companies aiming to launch Low Earth Orbit (LEO) satellite constellations to provide low-latency internet connectivity globally \cite{9502642}. This innovative approach is emerging as a promising alternative to traditional inter-continental underwater cables, which often fail to reach remote and underserved regions, such as isolated islands.
In this context, companies such as OneWeb, Viasat, Globalstar, and Amazon (Project Kuiper) are at the forefront, each striving to deploy vast networks of LEO satellites to ensure ubiquitous broadband coverage. SpaceX's Starlink project has demonstrated the feasibility and potential of this technology, spurring a competitive landscape where numerous other enterprises are rapidly advancing. These ambitious initiatives aim to bridge the digital divide, offering high-speed internet access to even the most remote corners of the world, thereby fostering greater global connectivity and inclusion.

\begin{figure}[t]
\centering
\centerline{\includegraphics[width=\columnwidth]{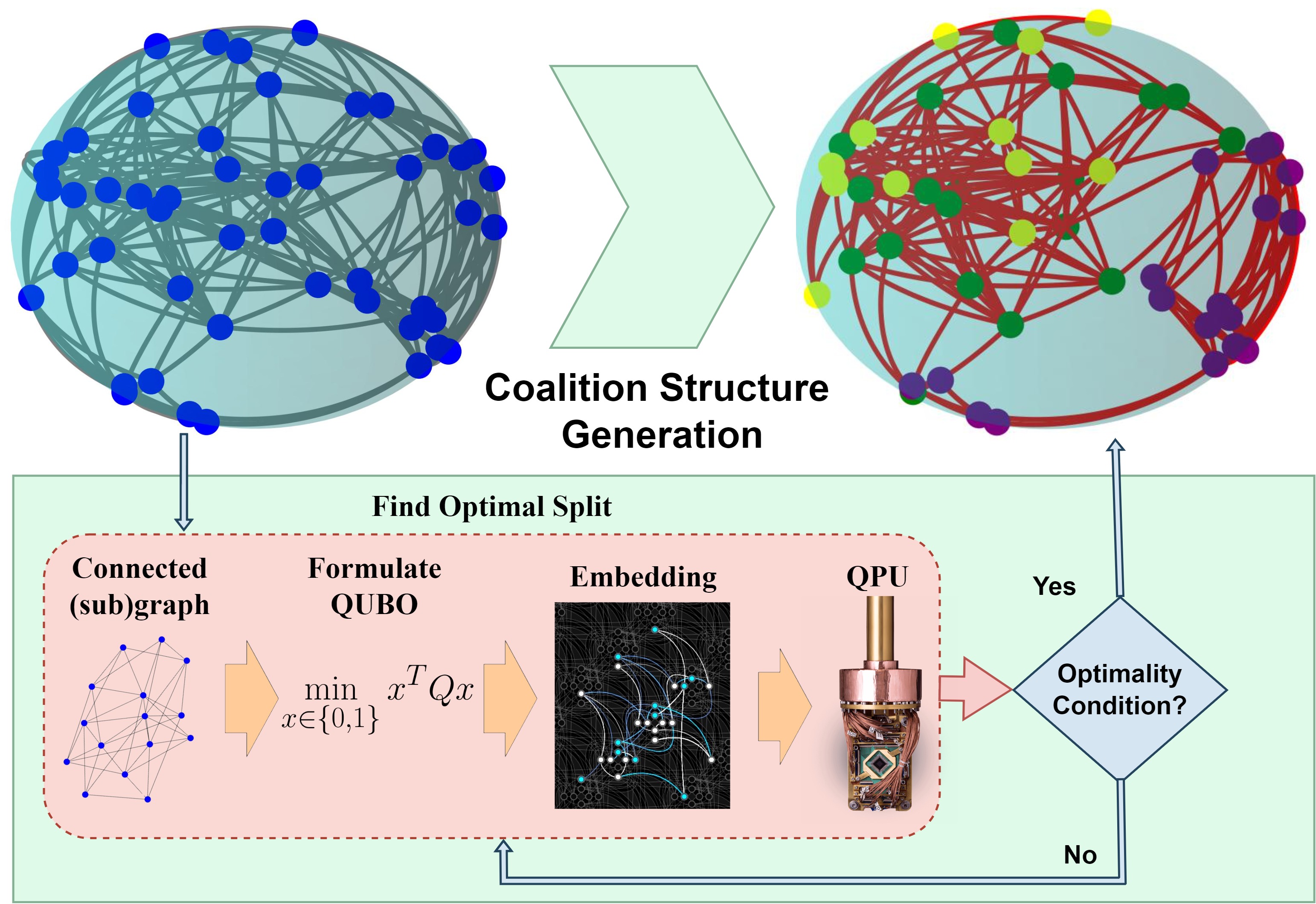}}
\caption{Architecture for finding the optimal coalition structures using \textit{GCS-Q}, an iterative algorithm for optimal graph partitioning leveraging quantum annealing hardware. At each step, the optimal graph cut is determined by formulating a QUBO problem, which is embedded onto the annealer hardware. The quantum tunneling phenomenon helps find the lowest energy solution efficiently within the exponentially large solution space.
The optimality condition involves choosing the best solution from the sample set output by the annealer that satisfies the use case-specific constraints.
}
\label{fig: poster}
\end{figure}

Despite the promising advantages of LEO satellite constellations, the peer-to-peer (P2P) communication model inherent to these networks presents significant challenges, particularly in terms of latency and network efficiency \cite{9210567}. In a P2P model, data packets must traverse multiple satellite nodes to reach their destination, which can result in increased latency due to the numerous hops required. This latency issue is exacerbated by the dynamic nature of satellite orbits and the varying distances between satellites \cite{9466942}. Adopting a coalition formation paradigm, where satellites are grouped into clusters, can mitigate these challenges by reducing the number of hops needed for data transmission. Within this paradigm, while P2P communication still occurs, cluster heads play a crucial role in managing intra-cluster communication and coordinating with other cluster heads for inter-cluster communication. This structure shortens communication paths, minimizes latency, and improves overall network performance \cite{9544067}. Finding the optimal partitioning, taking into account various factors for reliable communication, is an NP-Hard task \cite{drones8020044, 9544067}. With thousands of LEO satellites continuously in motion, the need to frequently find optimal coalitions adds to the sheer complexity of the task. There is a trade-off between experiencing higher latency due to more hops in the P2P communications without clusters and spending computational time to evaluate the optimal clusters.

In this paper, we formulate the clustering of LEO satellites as a coalition structure generation (CSG) problem and leverage the power of near-term quantum annealer hardware to solve it. We represent the topology of the LEO satellites as a graph and define the problem of finding the optimal partition of satellites into distinct coalitions by iteratively partitioning the graph. Given the exponential complexity of graph partitioning, we reformulate the problem as a Quadratic Unconstrained Binary Optimization (QUBO) problem, which can be efficiently addressed using quantum annealers \cite{10.1007/978-3-031-36030-5_11}.
We conduct experiments on simulations based on existing literature and explore various configurations to test different sparsity levels of communication links among the satellites. Utilizing real hardware by D-Wave, accessed remotely as a cloud service, we empirically compare the performance against the classical baseline solver Gurobi. Our results demonstrate a superior performance of the quantum annealer over the classical state-of-the-art method, highlighting the potential of quantum computing in addressing complex optimization problems in satellite communication networks.

\section{Related Works}
While increasing the number of satellites in a network enhances throughput, it also raises management overhead and inter-satellite interference, leading to only marginal improvements. To address this, the concept of satellite clusters has been explored to boost performance through cooperation \cite{8293791}. Typically, satellites are treated as autonomous agents equipped with sensors and computing capabilities to process requests, and they are constantly in motion. As satellites can communicate with one another, a cooperative game-theoretical framework can be employed to find the optimal partition of the satellites into mutually disjoint subsets that maximize utility. Specifically, the use case of satellites is formulated as a Graph Coalition Game, and solving this problem is NP-Hard.

Considering use-case-specific constraints such as cluster size, connectivity, and diameter, several works aim to optimize one or more network management factors, such as stable communications \cite{9544067, 8809684}, energy-efficient task cooperation \cite{10000401}, task allocation, and resource utilization \cite{9613796, 8431278, 9207866, 9165752}. Leveraging small groups of satellites to minimize routing overhead in wireless ad-hoc networks has also been explored \cite{6133939, 7279199, 7380539, 8058294}. Additionally, there are classical methods inspired by quantum evolutionary algorithms for finding the optimal coalitions of satellites \cite{MOUSAVI201926, 8406915}. Our approach leverages these foundations but utilizes quantum annealing for enhanced efficiency.

Further, \cite{Bass_2018} discusses the use of quantum annealers for grouping satellites into coalitions to maximize the total coverage of a designated Earth region. However, the algorithm is impractical as the number of qubits grows in the order of the Stirling number \cite{mansour2016commutation}. In contrast, we build upon the work of \cite{10.1007/978-3-031-36030-5_11} by applying \textit{GCS-Q}, a quantum annealing-based graph coalition structure generation algorithm to find the optimal partition of the satellites that can run on existing hardware. Moreover, our method demonstrates high scalability and outperforms classical state-of-the-art methods in terms of runtime while generating the similar quality results.

\section{Methods}

\subsection{Graph Representation of LEO Satellite Network}

The LEO satellite network is represented as a connected, undirected, weighted graph \( G = (V, E, w) \), where \( V \) is the set of vertices representing the satellites, \( E \) is the set of edges representing the communication links between satellites, and \( w: E \to \mathbb{R} \) is the weight function assigned to each edge. An edge \( (u, v) \in E \) exists if there is a direct communication link between satellites \( u \) and \( v \). The edge weight \( w(u, v) \) is a function of the factors that contribute to the efficiency of the communication link. Based on a comprehensive survey of existing literature \cite{10.5555/3635637.3662987, 10000401, drones8020044, 9544067}, edge weight can be defined as:

\begin{equation}
    w(u, v) = \alpha \cdot L(u, v) + \beta \cdot R(u, v) + \gamma \cdot M(u, v) + \delta \cdot B(u, v)
\end{equation}

where \( L(u, v) \) represents the latency, which is a function of the distance between the communicating satellites \( u \) and \( v \); \( R(u, v) \) denotes the reliability, which can be the probability that the communication link between \( u \) and \( v \) is operational; \( M(u, v) \) accounts for the management overhead or the cost of maintaining the link; and \( B(u, v) \) indicates the bandwidth of the communication link.

The coefficients \(\alpha, \beta, \gamma, \delta\) are weighting factors that balance the influence of each component on the overall edge weight. These parameters are set depending on the specific use case.

\begin{figure*}[htbp]
  \centering
  \includegraphics[width=\textwidth]{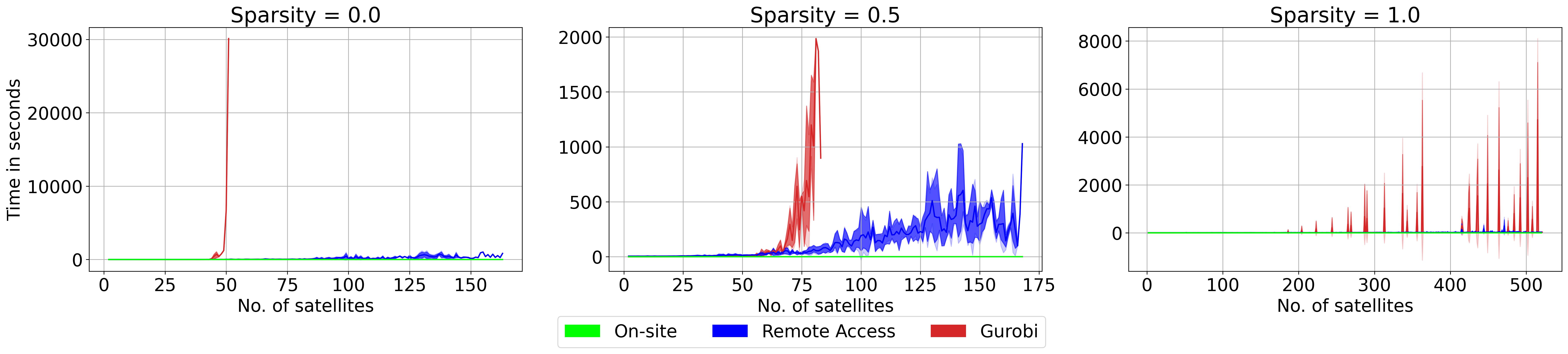}
  \vspace*{-16pt}
  \caption{Comparison of runtimes for the D-Wave Advantage annealer and Gurobi in finding the optimal graph partition for varying edge sparsities, where sparsity = 0 denotes a fully connected graph, and sparsity = 1 indicates the graph is a tree. The graph includes plots for the total runtime for remotely accessing the annealer as a cloud service, the runtime excluding internet latency and service queue waiting time (On-site), and the time taken for Gurobi to run locally. The plot illustrates the mean (represented by a solid line), the range (indicated by the broadly shaded area), and the standard deviation (denoted by the lightly shaded area) of the runtimes aggregated over three sets of synthetic data.}
  \vspace*{-16pt}
  \label{fig: runtime}
\end{figure*}

\subsection{Defining Graph Coalition Structure Generation Problem}

Given the graph $G(V,E,w)$ representing the satellite network, the graph coalition structure generation is an optimization problem that aims to find a partition ($P$) of $V$ that maximizes the total utility, which can be formulated as:

\begin{equation}\label{eq: CSG on ISG}
    P^* = \arg \max_{P} \sum_{C \in P} \sum_{i,j \in C} w(i,j)
\end{equation}

where $C \subseteq V$ is a coalition of satellites, and $P^*$ is the optimal coalition structure.
However, unlike continuous optimization, the cost function in this context is non-convex, meaning that small changes in input do not necessarily result in small changes in the cost. Additionally, the problem space contains an exponential number of potential solutions. To address these challenges, we leverage the quantum tunneling phenomenon, which allows us to efficiently navigate the solution space and identify the optimal solution.

\subsection{GCS-Q Algorithm}

\textit{GCS-Q} \cite{10.1007/978-3-031-36030-5_11} is a quantum-supported solution for finding the optimal coalition structure in graph games. Given the graph game, \textit{GCS-Q} initially assigns all agents to a single coalition and then follows a top-down approach by iteratively splitting the graph into two unconnected components. 

To find an optimal split (or optimal bipartition), \textit{GCS-Q} divides the underlying connected component of the graph (a coalition) into two disconnected subgraphs by removing the edges that maximize the sum of the remaining edge weights in the subgraphs. At each step, the quantum annealer is invoked to find the optimal split, exploring an exponential number of solutions. This leverages the power of the quantum annealing process, as finding the optimal bipartition of a connected graph is computationally expensive for classical computers.

Whenever a split of a coalition is found to have a better coalition value than the original coalition, it is considered. This process is repeated until no further beneficial splits are found. In our implementation, we use \textit{GCS-Q} for a constrained setting specific to the application, taking into account the maximum size of a cluster among the satellites.

\begin{figure*}[htbp]
  \centering
  \includegraphics[width=\textwidth]{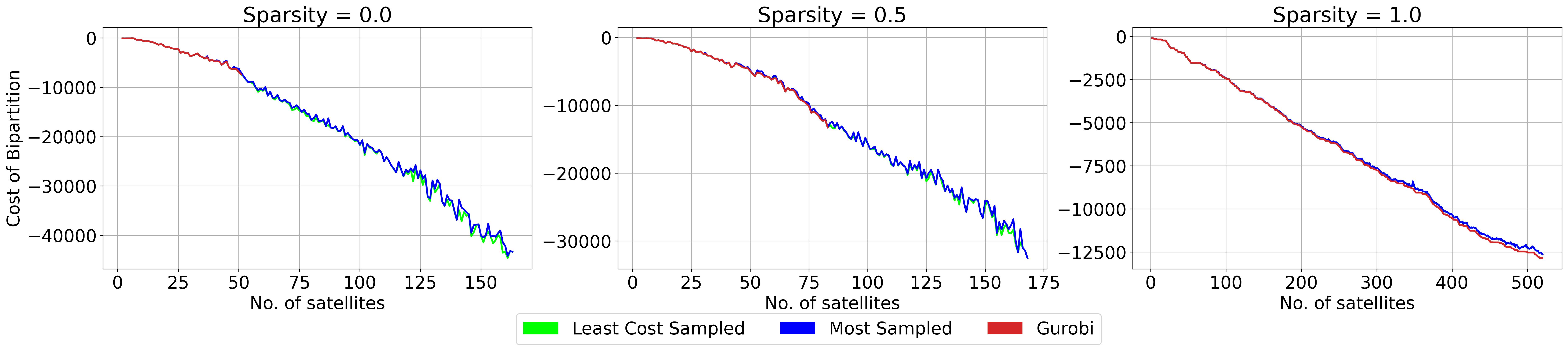}
  \vspace*{-16pt}
  \caption{Comparison of the solution quality for the single split problem across varying levels of graph sparsity. Sparsity = 0 denotes a fully connected graph, sparsity = 0.5 represents a graph with intermediate connectivity, and sparsity = 1 indicates a tree structure. The plots show the cost of bipartition against the number of satellites for one of the seeds, comparing the least cost sampled by the annealer, the most frequently sampled cost by the annealer, and the cost obtained by Gurobi.}
  \vspace*{-16pt}
  \label{fig: quality}
\end{figure*}

\section{Experiments}

\subsection{Experimental Settings}

We executed the experiments for \textit{GCS-Q} on the \textit{D-Wave Advantage} hardware provided as a cloud service via the \textit{dimod} library. To evaluate the performance of the existing hardware, we analyze both runtime and quality for single graph partitions as well as the complete coalition structure generation. As a classical counterpart, we implemented the state-of-the-art Gurobi solver and compare its performance with the annealer.
We executed the experiments on $3$ sets of synthetically generated data using fixed seeds for reproducibility and performed the runtime analysis.
The classical part of our experiments was conducted on a system equipped with a \textit{12th Gen Intel(R) Core(TM) i7-12800H} CPU @ \textit{2.40 GHz} and \textit{64 GB} RAM, with all software developed in \textit{Python 3.12}.

For real-world data, we captured the general perturbations (GP) orbital data of the satellites from Celestrak \footnote{https://celestrak.org/NORAD/elements/}. Specifically, we downloaded the three-line element (TLE/3LE) data of the Starlink satellites, and given a timestamp, we implemented code to evaluate the positions of the satellites. We ran the experiments limiting the number of satellites considered, starting with a small number and gradually increasing the number of satellites.
As the information about the communication links is not available, we construct a geometric graph \cite{pach2013beginnings} using a tunable radius parameter that defines the coverage for each satellite for P2P communication.

\subsection{Results}

In this section, we present the outcomes of our experiments, focusing on two main aspects: the optimal split of the satellite network graph and the application of real-world Starlink data. The results highlight the performance and quality of solutions obtained using the D-Wave Advantage annealer compared to the Gurobi solver.

\subsubsection{Optimal Split}

The algorithm iteratively solves for finding the optimal split in each step, which is also NP-Hard. 
Thus, we initially compare the performances of the Advantage annealer and Gurobi on graph partitioning for graphs that have varying levels of connectivity among the nodes. Specifically, we generated graphs that are fully connected (i.e., for $n$ nodes, there are $n(n-1)/2$ edges, considered as sparsity = 0) to a tree (i.e., for $n$ nodes, there are $n-1$ edges, considered as sparsity = 1). The 

\paragraph{Runtime}

We ran the annealer experiments by sending the problem from the client machine via the internet, with the annealer hardware accessed as a cloud service. It is important to note that the device is not dedicated exclusively to our experiments; thus, the problem incurs internet latency as it is sent as an HTTP request, waits in a service queue for our turn, and then the QPU access is provided for the problem. After the problem is solved, the sample set is sent back to our client machine as an HTTP response, which is parsed to find the optimal solution. We compare the runtime for this entire process against that of Gurobi, which runs locally on the client machine.

It is also shown in the literature that internet latency and the waiting time in the service queue contribute significantly to the total runtime \cite{venkatesh2023q}. However, for a fair comparison, we should assume a hypothetical scenario of having the annealer hardware on-site, eliminating the internet latency and service waiting time. In that case, we can only consider the QPU access time and the runtimes of the remaining classical operations in the algorithm \cite{Cellini2024}.

The results in Figure \ref{fig: runtime} indicate the efficiency of the annealer's runtime. Even the total runtime of the remotely accessed annealer is less than that of Gurobi executed locally for large dense graphs. Additionally, the runtime for the hypothetical scenario of having an on-site annealer is in the order of less than a second, as our subscription to D-Wave is capped at $1$ second, appearing flat compared to Gurobi, which typically took several seconds. 
We ran the experiments by increasing the number of graph nodes on the annealer till we reached the hardware limitation (see x-axis in Fig. \ref{fig: runtime} and  \ref{fig: quality}).
In contrast, we couldn't continue the experiments for Gurobi approximately beyond the size $50$ for sparsity $0$, size $80$ for sparsity $0.5$, and size $520$ for sparsity $1$ as it was allowed to run even for days but still couldn't output a solution.
There are also cases where it took a very long time or even failed to find a solution for certain problem instances when the branch and bound technique followed by Gurobi could not prune any portion of the exponentially large solution space and ended up searching it entirely.
Theoretically, the annealer takes constant time to find the lowest energy state. However, embedding and sampling multiple times are necessary as annealers are probabilistic machines.

\paragraph{Quality}

Although Gurobi is a state-of-the-art solver for combinatorial optimization, it is still an approximate solver like an annealer. While \textit{D-Wave} exhibits faster processing than \textit{Gurobi}, it is essential to evaluate the quality of the solutions. Figure \ref{fig: quality} shows an empirical analysis of the annealer's performance in terms of the quality of the solution obtained for the graph partition based on the value of the cut. The annealer consistently samples a solution that is as good as the one output by Gurobi.

Furthermore, as the annealer works as a sampling device expected to sample the optimal solution, the current implementation of D-Wave considers the lowest-valued sample as the solution. Ideally, an annealer device should consider the most frequently occurring sample as the solution. As we can set the number of times to sample, out of the $1000$ samples we took for each problem, we also plot the value of the most frequently occurring sample. The present-day performance of the D-Wave Advantage is impressive, rendering the difference negligible. This indicates that the problem being solved is well-embeddable onto the annealer hardware, which aids in faster processing and maintaining high quality.

\subsubsection{Starlink Data}

By using the TLE/3LE data, we obtain the satellite positions for a specific timestamp and constructed a geometric graph. The edge weights were assigned based on the distance, adding a random interference noise component, such that both positive and negative values were allowed.

Implementing the constraint on the maximum number of satellites in a coalition is a straightforward modification for the annealers. In a scenario where a coalition has no splits with a total value greater than that of the coalition under consideration but exceeds the expected number of satellites, the annealer can choose the next best solution present in the sample set without mathematically modifying the problem formulation. In contrast, for implementing the constraint on the cluster size using Gurobi, the QUBO problem needs to explicitly incorporate this constraint.

For our experiments, we set a constant timestamp and calculated the positions of the satellites. By setting a coverage parameter, we constructed the graph with edge weights as a function of distance and an interference noise drawn from a random distribution. We also set the maximum coalition size to $5$.
Since every communication link has a management cost associated with it \cite{9613796}, we compared the usefulness of coalition formation by evaluating the number of communication links before and after coalition formation. Our results indicate a significant benefit from clustering the satellites, which simplifies management compared to handling all satellites individually.

\begin{figure}[t]
\centering
\centerline{\includegraphics[width=\columnwidth]{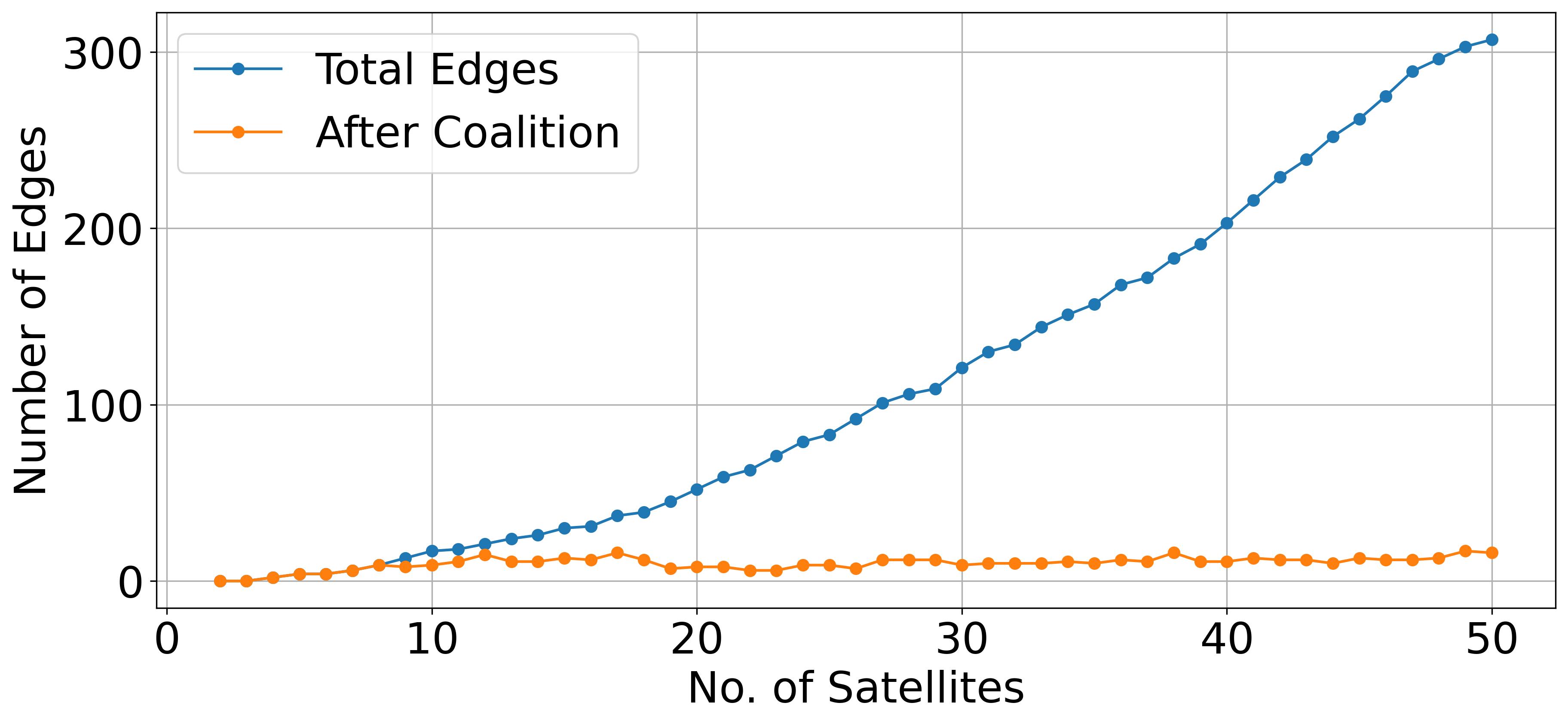}}
\caption{The figure depicts the benefits of coalition formation in terms of the number of communication links. Having a lower number of communication links is preferable as each link has a management cost.}
\label{fig: links}
\end{figure}

\section{Discussion}

In this paper, we proposed a quantum annealing-based approach to address the coalition structure generation problem for Low Earth Orbit (LEO) satellite networks. 
Applying \textit{GCS-Q}, the D-Wave Advantage demonstrated superior runtime performance compared to the classical Gurobi solver. Even when accessed remotely, the annealer's total runtime was less than that of Gurobi for dense graphs, which ran locally. This efficiency becomes even more pronounced when considering an on-site annealer scenario, eliminating internet latency and service queue waiting times.
The quality of the solutions obtained from the quantum annealer was comparable to those from Gurobi. The annealer consistently found solutions with optimal or near-optimal values, highlighting its effectiveness in solving complex combinatorial optimization problems.

With a simple change in the implementation for considering the solution from the sample set of the annealer, \textit{GCS-Q} showed the ability to handle varying levels of graph connectivity and constraints, such as maximum coalition size, underscoring its scalability and practicality for real-world applications. Our experiments with Starlink satellite data further validated the approach, showing that the method can adapt to dynamic and large-scale satellite networks.
The proposed hybrid quantum-classical approach offers significant potential for enhancing the management of LEO satellite networks. By optimizing coalition structures, the network can achieve reduced latency, improved reliability, and more efficient resource utilization.
Moreover, the number of requests each satellite can send and receive is typically limited. For instance, if a satellite has an antenna capable of sending a message to 5 nearby satellites simultaneously, in terms of the graph problem formulated, the degree of a vertex should be a maximum of 5. While this specific constraint has not been explored in our current study, it is noteworthy that with the existing hardware capabilities and the ability to tweak the chain strength parameter of the D-Wave annealer, one could potentially adjust this parameter to control the chain break fraction. This adjustment could effectively translate to the constraint on the degree of a vertex, representing a promising direction for future research.

\section{Conclusion and Future Work}

The integration of quantum computing into satellite network management represents a promising frontier. Our study illustrates that quantum annealing can effectively tackle the NP-Hard problem of coalition structure generation in LEO satellite networks, outperforming classical methods in runtime and maintaining high solution quality. As quantum hardware continues to advance, we anticipate even greater improvements in the optimization of complex networks.

Future work will focus on leveraging the limitation of the present-day hardware to tackle the constraint on the degree of each vertex by tuning the chain strength.
Also, finding the optimal splits can be parallelized as each step would be working on unconnected components of the graph.
As a classical solver like Gurobi can work well on sparse graphs, one can implement a condition for deciding whether to use Annealer or classical methods at each step.
Moreover, with the availability of dynamic proprietary information about the satellites' communication links and management costs, empirical advantages can be proven more evidently.
With the advent of gate-based quantum computers, variational quantum algorithms incorporating qubit-efficient encoding strategies can be employed to solve the QUBO problem as an alternative for annealers \cite{venkatesh2024qubit}.
This research lays a foundation for leveraging quantum computing to address other challenging optimization problems in the satellite domain, paving the way for more robust and efficient network infrastructures.

\section*{Code Availability}
All code to generate the data, figures, analyses, as well as, additional technical details on the experiments are publicly available at \href{https://github.com/supreethmv/LEO-satellites-coalition}{https://github.com/supreethmv/LEO-satellites-coalition}.

\section*{Acknowledgment}
S. Mysore Venkatesh acknowledges support through a scholarship from the University of Kaiserslautern-Landau.

\bibliographystyle{IEEEtran}
\bibliography{references}

\begin{thebibliography}{10}
\providecommand{\url}[1]{#1}
\csname url@samestyle\endcsname
\providecommand{\newblock}{\relax}
\providecommand{\bibinfo}[2]{#2}
\providecommand{\BIBentrySTDinterwordspacing}{\spaceskip=0pt\relax}
\providecommand{\BIBentryALTinterwordstretchfactor}{4}
\providecommand{\BIBentryALTinterwordspacing}{\spaceskip=\fontdimen2\font plus
\BIBentryALTinterwordstretchfactor\fontdimen3\font minus \fontdimen4\font\relax}
\providecommand{\BIBforeignlanguage}[2]{{%
\expandafter\ifx\csname l@#1\endcsname\relax
\typeout{** WARNING: IEEEtran.bst: No hyphenation pattern has been}%
\typeout{** loaded for the language `#1'. Using the pattern for}%
\typeout{** the default language instead.}%
\else
\language=\csname l@#1\endcsname
\fi
#2}}
\providecommand{\BIBdecl}{\relax}
\BIBdecl

\bibitem{9502642}
S.~Liu, Z.~Gao, Y.~Wu, D.~W. Kwan~Ng, X.~Gao, K.-K. Wong, S.~Chatzinotas, and B.~Ottersten, ``Leo satellite constellations for 5g and beyond: How will they reshape vertical domains?'' \emph{IEEE Communications Magazine}, vol.~59, no.~7, pp. 30--36, 2021.

\bibitem{9210567}
O.~Kodheli, E.~Lagunas, N.~Maturo, S.~K. Sharma, B.~Shankar, J.~F.~M. Montoya, J.~C.~M. Duncan, D.~Spano, S.~Chatzinotas, S.~Kisseleff, J.~Querol, L.~Lei, T.~X. Vu, and G.~Goussetis, ``Satellite communications in the new space era: A survey and future challenges,'' \emph{IEEE Communications Surveys \& Tutorials}, vol.~23, no.~1, pp. 70--109, 2021.

\bibitem{9466942}
N.~Saeed, H.~Almorad, H.~Dahrouj, T.~Y. Al-Naffouri, J.~S. Shamma, and M.-S. Alouini, ``Point-to-point communication in integrated satellite-aerial 6g networks: State-of-the-art and future challenges,'' \emph{IEEE Open Journal of the Communications Society}, vol.~2, pp. 1505--1525, 2021.

\bibitem{9544067}
J.~Liu, X.~Zhang, R.~Zhang, T.~Huang, and F.~R. Yu, ``Reliable and low-overhead clustering in leo small satellite networks,'' \emph{IEEE Internet of Things Journal}, vol.~9, no.~16, pp. 14\,844--14\,856, 2022.

\bibitem{drones8020044}
M.~Zhuo, Y.~Feng, P.~Yang, Z.~Tian, L.~Liu, and S.~Zhou, ``Optimizing topology in satellite–uav collaborative iot: A graph partitioning simulated annealing approach,'' \emph{Drones}, vol.~8, no.~2, 2024.

\bibitem{10.1007/978-3-031-36030-5_11}
S.~M. Venkatesh, A.~Macaluso, and M.~Klusch, ``Gcs-q: Quantum graph coalition structure generation,'' in \emph{Computational Science -- ICCS 2023}, J.~Miky{\v{s}}ka, C.~de~Mulatier, M.~Paszynski, V.~V. Krzhizhanovskaya, J.~J. Dongarra, and P.~M. Sloot, Eds.\hskip 1em plus 0.5em minus 0.4em\relax Cham: Springer Nature Switzerland, 2023, pp. 138--152.

\bibitem{8293791}
G.-P. Liu and S.~Zhang, ``A survey on formation control of small satellites,'' \emph{Proceedings of the IEEE}, vol. 106, no.~3, pp. 440--457, 2018.

\bibitem{8809684}
N.~Xing, Q.~Zong, L.~Dou, B.~Tian, and Q.~Wang, ``A game theoretic approach for mobility prediction clustering in unmanned aerial vehicle networks,'' \emph{IEEE Transactions on Vehicular Technology}, vol.~68, no.~10, pp. 9963--9973, 2019.

\bibitem{10000401}
Y.~Huang, N.~Qi, Z.~Huang, L.~Jia, Q.~Wu, R.~Yao, and W.~Wang, ``Connectivity guarantee within uav cluster: A graph coalition formation game approach,'' \emph{IEEE Open Journal of the Communications Society}, vol.~4, pp. 79--90, 2023.

\bibitem{9613796}
Z.~Gao, A.~Liu, C.~Han, and X.~Liang, ``Files delivery and share optimization in leo satellite-terrestrial integrated networks: A noma based coalition formation game approach,'' \emph{IEEE Transactions on Vehicular Technology}, vol.~71, no.~1, pp. 831--843, 2022.

\bibitem{8431278}
F.~Afghah, M.~Zaeri-Amirani, A.~Razi, J.~Chakareski, and E.~Bentley, ``A coalition formation approach to coordinated task allocation in heterogeneous uav networks,'' in \emph{2018 Annual American Control Conference (ACC)}, 2018, pp. 5968--5975.

\bibitem{9207866}
J.~Chen, Q.~Wu, Y.~Xu, N.~Qi, X.~Guan, Y.~Zhang, and Z.~Xue, ``Joint task assignment and spectrum allocation in heterogeneous uav communication networks: A coalition formation game-theoretic approach,'' \emph{IEEE Transactions on Wireless Communications}, vol.~20, no.~1, pp. 440--452, 2021.

\bibitem{9165752}
H.~Luan, Y.~Xu, D.~Liu, Z.~Du, H.~Qian, X.~Liu, and X.~Tong, ``Energy efficient task cooperation for multi-uav networks: A coalition formation game approach,'' \emph{IEEE Access}, vol.~8, pp. 149\,372--149\,384, 2020.

\bibitem{6133939}
K.~Abboud and W.~Zhuang, ``Impact of node clustering on routing overhead in wireless networks,'' in \emph{2011 IEEE Global Telecommunications Conference - GLOBECOM 2011}, 2011, pp. 1--5.

\bibitem{7279199}
------, ``Impact of microscopic vehicle mobility on cluster-based routing overhead in vanets,'' \emph{IEEE Transactions on Vehicular Technology}, vol.~64, no.~12, pp. 5493--5502, 2015.

\bibitem{7380539}
P.~Thakur and A.~Ganpati, ``Cluster based route discovery technique for routing protocol in manet,'' in \emph{2015 International Conference on Green Computing and Internet of Things (ICGCIoT)}, 2015, pp. 622--626.

\bibitem{8058294}
S.~Shruthi, ``Proactive routing protocols for a manet — a review,'' in \emph{2017 International Conference on I-SMAC (IoT in Social, Mobile, Analytics and Cloud) (I-SMAC)}, 2017, pp. 821--827.

\bibitem{MOUSAVI201926}
\BIBentryALTinterwordspacing
S.~Mousavi, F.~Afghah, J.~D. Ashdown, and K.~Turck, ``Use of a quantum genetic algorithm for coalition formation in large-scale uav networks,'' \emph{Ad Hoc Networks}, vol.~87, pp. 26--36, 2019. [Online]. Available: \url{https://www.sciencedirect.com/science/article/pii/S1570870518303044}
\BIBentrySTDinterwordspacing

\bibitem{8406915}
------, ``Leader-follower based coalition formation in large-scale uav networks, a quantum evolutionary approach,'' in \emph{IEEE INFOCOM 2018 - IEEE Conference on Computer Communications Workshops (INFOCOM WKSHPS)}, 2018, pp. 882--887.

\bibitem{Bass_2018}
\BIBentryALTinterwordspacing
G.~Bass, C.~Tomlin, V.~Kumar, P.~Rihaczek, and J.~Dulny, ``Heterogeneous quantum computing for satellite constellation optimization: solving the weighted k-clique problem,'' \emph{Quantum Science and Technology}, vol.~3, no.~2, p. 024010, mar 2018. [Online]. Available: \url{https://dx.doi.org/10.1088/2058-9565/aaadc2}
\BIBentrySTDinterwordspacing

\bibitem{mansour2016commutation}
T.~Mansour and M.~Schork, \emph{Commutation relations, normal ordering, and Stirling numbers}.\hskip 1em plus 0.5em minus 0.4em\relax CRC Press Boca Raton, 2016, vol.~2.

\bibitem{10.5555/3635637.3662987}
X.~Lu, H.~Song, H.~Ma, and J.~Zhu, ``A task-driven multi-uav coalition formation mechanism,'' in \emph{Proceedings of the 23rd International Conference on Autonomous Agents and Multiagent Systems}, ser. AAMAS '24.\hskip 1em plus 0.5em minus 0.4em\relax Richland, SC: International Foundation for Autonomous Agents and Multiagent Systems, 2024, p. 1292–1300.

\bibitem{pach2013beginnings}
J.~Pach, ``The beginnings of geometric graph theory,'' in \emph{Erd{\H{o}}s Centennial}.\hskip 1em plus 0.5em minus 0.4em\relax Springer, 2013, pp. 465--484.

\bibitem{venkatesh2023q}
\BIBentryALTinterwordspacing
S.~M. Venkatesh, A.~Macaluso, M.~Nuske, M.~Klusch, and A.~Dengel, ``Q-seg: Quantum annealing-based unsupervised image segmentation,'' 2023. [Online]. Available: \url{https://arxiv.org/abs/2311.12912}
\BIBentrySTDinterwordspacing

\bibitem{Cellini2024}
\BIBentryALTinterwordspacing
L.~Cellini, A.~Macaluso, and M.~Lombardi, ``Qal-bp: an augmented lagrangian quantum approach for bin packing,'' \emph{Scientific Reports}, vol.~14, no.~1, p. 5142, Mar 2024. [Online]. Available: \url{https://doi.org/10.1038/s41598-023-50540-3}
\BIBentrySTDinterwordspacing

\bibitem{venkatesh2024qubit}
S.~M. Venkatesh, A.~Macaluso, M.~Nuske, M.~Klusch, and A.~Dengel, ``Qubit-efficient variational quantum algorithms for image segmentation,'' \emph{arXiv preprint arXiv:2405.14405}, 2024.

\end{thebibliography}

\end{document}